\documentclass[english]{article}
\usepackage{geometry}
\usepackage[T1]{fontenc}
\usepackage[latin1]{inputenc}
\makeatletter
\usepackage{babel}
\makeatother
\begin{document}

\title{Origin of the pearl necklace of SN1987A}

\author{Jacques Moret-Bailly}

\maketitle
e-mail: jacques.moret-bailly@u-bourgogne.fr

\begin{abstract}
The bright circles observed around stars are usually considered as produced by shock waves; but this interpretation does not explain easily the bright spots of the \textquotedblleft pearl necklace \textquotedblright of NS 1987A supernova.

Assuming that the central object of SN 1987A is a neutron star heated by the accretion of a low density cloud, non-linear optics, in particular superradiance and impulsive stimulated Raman scattering (ISRS), is needed to take into account the high intensity of the radiated light.

Where the temperature of the surrounding gas decreases enough to allow a combination of protons and electrons into atomic hydrogen in despite of the low density, a spherical shell absorbs in particular the Lyman alpha line, but does not populate much the 2P state because a tangential superradiance appears until the exciting line is almost absorbed; the increase of the 2P population resulting from the disappearance of the superradiance produces a redshift, so that almost all energy of a wide band is transferred to tangential modes making an UV pearl necklace in a given direction of observation.

In a column of UV light making a pearl, atomic lines are excited enough to produce new, co-linear superradiances, in particular visible.
\end{abstract}
\section{Introduction}
Supernova SN1987A shows a ''pearl necklace'' and infrared emissions whose interpretation appears difficult (Bouchet et al. 2006, Lawrence et al. 2000, Thornhill \& Ransom 2006). A source or these particularities may be simply spectroscopy.

Assuming that the central object of SN 1987A is a neutron star heated by the accretion of a low density hydrogen cloud, the density of emitted energy is high, so that the transfers of energy are non-linear, strong. Such effects are usually observed with lasers.

Section 2, recalls some of these effects, specifying the language and the notations.

In section 3 the generation of the pearl necklace appears as a trivial application of these effects.
  
\section{Some high energy optical effects.}
\subsection{Superradiant emissions.}
Planck used the notion of electromagnetic modes introduced in the nineteenth century to study music instruments. For this application, the used modes were restricted to stationary systems, while the mathematical definition is wider: A mode is a ray in the real vector space of the solutions of a linear set of equations (that is, all solutions in a mode depend only on a single real parameter named amplitude of the solution). Happily, the  definition of the modes does not imply stationary and monochromatic conditions, because such conditions would require an infinite duration of the experiments.

The electromagnetic fields in the vacuum obey the linear Maxwell's equations up to very high frequencies. Schwarzschild and Fokker trick allows  introducing sources replaced by their advanced fields.

Planck mistook computing the additive term in his equation which gives the energy of an electromagnetic mode inside a blackbody at temperature $T$, defining also the temperature of a mode. Nernst (1916) found the right value $h\nu/2$, so that the correct mean energy in a monochromatic mode of frequence $\nu$ writes $h\nu(1/(\exp(h\nu/kT)-1)+1/2)$.

Einstein (1917) introduced the stimulated emission, but justifying the spontaneous emission, he did not remark that the electromagnetic field has a minimal mean value corresponding to $h\nu/2$ in a blackbody at 0K; laser experiments show that the spontaneous emission is a stimulated emission corresponding to the amplification of this minimal value.

Suppose that only two molecular levels of energies $e_1$ and $e_2$ ($e_2>e_1$) are implied in a transition of energy $\nu = (e_2-e_1)/h$. The molecular temperature $T_{12}$ relative to this transition is deduced from the molecular populations $n_1$ and $n_2$ in these levels by $n_1/n_2 = \exp(h\nu/kT_{12})$.

Suppose that a long cell is filled with this gas; a low temperature light beam entering into the cell increases the entropy of the system, being amplified, heated to reach temperature $T_{12}$. At the beginning, the amplification may be said a spontaneous emission, then it is induced, the energy of the beam increasing proportionally to this energy, that is exponentially. But the available molecular energy is limited, so that the temperature $T_{12}$ {} is decreased, usually strongly. Thus, spontaneous emissions in other directions are strongly decreased. This decrease resulting from a large excitation seems a paradox.

If a long path in a strongly excited medium is real, it is a superradiant emission; if it is virtual, using mirrors, it is a laser emission.

\subsection{The \textquotedblleft Impulsive Stimulated Raman Scattering (ISRS).\textquotedblright }

It is usually assumed that there is no interaction between light beams refracted by a transparent medium. Experiments and a regular theory (Giordmaine et al. (1968), Yan et al. (1985), Weiner et al. (1990), Dougherty et al. (1992), Dhar et al. (1994), ...) show that this assumption is wrong if light and matter verify conditions set by Lamb (1971). The interaction which increases the entropy of a set of refracted usual time-incoherent beams by frequency shifts, without blurring the images and the spectra, works well in atomic hydrogen in its first excited state. Planck's law and thermodynamics show that, usually, light is redshifted, and the energy it looses is transferred to the thermal background. There is no energy threshold of the energy of the beams, the ISRS becoming the Coherent Raman Effect on Incoherent Light (CREIL) at low intensities (Moret-Bailly 2005-6).

\section{Generation of the pearl necklace.}
\subsection{Absorption of light emitted by the neutron star.}
Accreting hydrogen, the neutron star becomes extremely hot, at least at hot spots; thus it emits light mainly in the far ultraviolet.

Close to the star, an atom is ionized; hydrogen and light element impurities lose all their electrons, producing a transparent plasma. Some heavier atoms may absorb light, but they re-emit energy in the UV, so that few energy is lost.

Assuming a low density, a combination of protons and electrons requires a low temperature, say 20 000 K. Hydrogen is mainly produced in its ground state, but it absorbs Lyman lines, so that a lot of its states may be populated.

Hydrogen absorbs frequencies lower than Lyman limit only on the its spectral lines. In next subsection, we will see that a superradiant emission decreases the population of the excited states, so that few 2P hydrogen remains, able to shift the spectrum, until, paradoxically, the Lyman $\alpha$ line is almost absorbed: When this happens, the superradiance decreases, so that more 2P hydrogen remains and shifts the light, the energy at the 2P frequency and at the frequencies of other absorption lines is renewed.

Finally, the absorption at a Lyman line, which is strong, nearly complete, sweeps a wide band.

\medskip
The final result is an extremely strong absorption not only at the frequencies of the lines (eventually not Lyman), but on wide bands corresponding to a redshift which lasts as long as energy is shifted to the Lyman $\alpha$ frequency. The energy lost by the redshifts, blueshifts the thermal background, that is heats it, simulating hot dust.

\subsection{Generation of an UV necklace.}
Previous subsection showed that the gas is strongly excited in a spherical shell centred on the neutron star, by a wide band absorption. Superradiant emissions may appear in directions for which the gas is thick, that is tangentially to the spheres. This system is very similar to a laser pumped transversally by a light-emitting diode of the same frequency. Although the gas is excited to eigenstates, the light emitted by the shell of excited gas makes a continuous spectrum because the column density of 2S hydrogen crossed from the emission point to outside is variable, producing variable redshifts. The frequency shift of the emitted lines prevents an absorption at the exciting (Lyman $\alpha$ ...) frequency.

Consider two thin, close spherical shells emitting lines whose local difference of frequencies is of the order of the linewidth; each line induces the emission of the other, so that their modes are bound; this binding extends to the whole superradiant band whose modes are, consequently, connected.

The strong absorptions and re-emissions in the shell of excited atomic hydrogen generate for each direction a bright ring in the UV. In this ring, the competition of the polychromatic modes selects a particular set of modes, producing, in the UV, bright spots. A nearly complete transfer of energy from the radial beams to the rings obeys thermodynamics because the solid angle on which a ring is seen is much larger than the solid angle on which the central engine could be seen.

\subsection{Generation of a visible necklace.}
Around the shell of far UV absorbing hydrogen, the columns of UV light emitted to the Earth excite hydrogen and various atoms strongly, so that, in the direction of the Earth, the emissions of these atoms are superradiant and co-linear to the columns: the visible pearl necklace is generated.

\section{Conclusion}
The present paper neglects a large part of the complexity of the problen: the necklace is not a circle; can the other circles be generated by a similar process around two other stars ? If yes, why did they illuminate when the necklace appeared ?

However, although partial, the present theory gives results using few hypothesis and regular physics. The explanation of the pearl necklace of SN1987A is rough, but it requires very few astrophysical hypothesis, its originality being the use of optical and spectroscopic properties more generally developed and well verified in laser technology. The visible necklace should come with a remaining UV spectrum in which absorptions corresponding to the lines observed in the visible may appear. Among other verifications, it could be a positive test of the present theory.

\section{Bibliography.}

  Bouchet, P., E. Dwek, I. J. Danziger, R. G. Arendt, I. J. M. De Buizer, S. Park, N. B. Suntzeff, R. P. Kirshner, P. Challis, 2006, arxiv:astro-ph/0601495

Dhar, L. , J. A. Rogers, \& K. A. Nelson, 1994 {\it Chem. Rev.} {\bf 94}, 157
 
Dougherty, T. P., G. P. Wiederrecht, K. A. Nelson, M. H. Garrett, H. P. Jenssen \& C. Warde,1992, {\it Science} {\bf 258,}, 770

Einstein A., 1917, {\it Phys. Z.}, {\bf 18} 121

Giordmaine, J. A. , M. A. Duguay \& J. W.Hansen, 1968, {\it IEEE J. Quantum Electron.}, {\bf }4,252

Lawrence S. S.,B. E. Sugerman, P. Bouchet, A. P. S. Crotts, R. Uglesich \& S. Heathcote, 2000, arxiv:astro-ph/0004191

Moret-Bailly, J., 2005, arxiv: physics/0507141

Moret-Bailly, J., 2006,  {\it AIP Conference Proceedings}, 822, 226-238

Nernst W. , {\it Verh. Deutsch. Phys. Ges}, {\bf 18}, 83 (1916)

Thornhill,W. W., \& C. J. Ransom 2006 {\it 2006 IEEE International Conference on Plasma Science}, 369

Weiner, A. M. , D. E. Leaird., G. P. Wiederrecht, \& K. A. Nelson,  1990, {\it Science}  {\bf 247}, 1317

Yan, Y.-X. , E. B. Gamble Jr. \& K. A. Nelson , 1985,  {\it J. Chem Phys.}, {\bf 83}, 5391

\end{document}